\documentclass{appolb}
\usepackage{epsfig}
\usepackage{amsmath}
\usepackage{amsfonts}
\usepackage{amssymb}
\usepackage{subfig}

\newcommand{\be}{\begin{equation}}
\newcommand{\ee}{\end{equation}}
\newcommand{\bes}{\begin{subequations}}
\newcommand{\ees}{\end{subequations}}
\newcommand{\bea}{\begin{eqnarray}}
\newcommand{\eea}{\end{eqnarray}}
\newcommand{\bear}{\begin{equation}\begin{array}}
\newcommand{\eear}[1]{\end{array}\label{#1}\end{equation}}
\def\ba{$$\begin{array}}
\def\ea{\end{array}$$}
\def\bra{$\begin{array}}
 \def\era{\end{array}$}

\begin{document}
\title{Dark matter data and quartic self-couplings in Inert Doublet Model%
\thanks{Presented at XXXV International Conference of Theoretical Physics MATTER TO THE DEEPEST: Recent Developments in Physics of Fundamental Interactions, Ustro\'n'11}%
}
\author{Dorota Soko\l owska
\address{University of Warsaw, Faculty of Physics}
}
\maketitle
\begin{abstract}
We analyse the thermal evolution of the Universe in the Inert Doublet Model for three viable regions of Dark Matter mass: low, medium and high DM mass. Those three regions exhibit different behaviour in the possible types of evolution. We argue that the quartic self-couplings in IDM are significant parameters for the astrophysical analysis. 
\end{abstract}
\PACS{12.60.Fr, 95.35.+d}
  
\section{Thermal evolution of the Universe in IDM}

The Inert Doublet Model (IDM) \cite{Deshpande:1977rw,Barbieri:2006dq} is a $Z_2$-symmetric 2HDM, which for a special set of parameters may provide the Dark Matter (DM) candidate.  The model contains two scalar $SU(2)$ doublets: a "standard" scalar (Higgs) doublet $\Phi_S$ and a "dark" scalar doublet $\Phi_D$. $\Phi_S$ is responsible for the electroweak symmetry breaking and masses of fermions and gauge bosons as in the Standard Model (SM), while $\Phi_D$ does not receive  vacuum expectation value (v.e.v.) and does not couple to fermions. In the model the discrete $D$-symmetry of the $Z_2$ type is present:
\begin{equation}
D: \quad \Phi_S \xrightarrow{D} \Phi_S,\quad
	\Phi_D \xrightarrow{D} -\Phi_D,\quad
	SM \textrm{ fields}     \xrightarrow{D} SM \textrm{ fields}.
	\label{dtransf}
\end{equation}

All the degrees of freedom of the dark doublet $\Phi_D$ are realized
as the massive  $D$-scalars: two charged $D^\pm$ and two neutral
$D_H$ and $D_A$. They possess a conserved multiplicative quantum number, \textit{the odd $D$-parity},
and therefore the lightest particle among them can be considered as a candidate for the DM particle.

The $D$-symmetric potential $V$, which can describe IDM, is:
\begin{eqnarray}
V = -\frac{m_{11}^2}{2} \Phi_S^\dagger\Phi_S - \frac{m_{22}^2}{2} \Phi_D^\dagger\Phi_D + \frac{\lambda_1}{2} \left(\Phi_S^\dagger\Phi_S\right)^2 + \frac{\lambda_2}{2} \left(\Phi_D^\dagger\Phi_D \right)^2 + \nonumber \\
\lambda_3 \left(\Phi_S^\dagger\Phi_S \right) \left(\Phi_D^\dagger\Phi_D\right) + \lambda_4 \left(\Phi_S^\dagger\Phi_D\right) \left(\Phi_D^\dagger\Phi_S\right) +\frac{\lambda_5}{2}\left[\left(\Phi_S^\dagger\Phi_D\right)^2\!+\!h.c. \right], \label{pot}
\end{eqnarray}
with all parameters real and $\lambda_5 <0$ \cite{Ginzburg:2010wa}. 
\textit{Positivity conditions} imposed on the potential guarantee that the extremum with the lowest energy will be the global minimum of the potential (vacuum). Relevant conditions are:
$ \lambda_{1,2} >0 \,,\; R+1 >0 \,; \, R = \lambda_{345}/\sqrt{\lambda_1 \lambda_2}\,,\; \lambda_{345} = \lambda_3 + \lambda_4 + \lambda_5$.

The Yukawa interaction of SM fermions $\psi_f$
with only one scalar doublet $\Phi_S$ have the same form as in the SM with the change $\Phi\to\Phi_S$ (Model I for a general 2HDM).

We consider \textit{thermal evolution of the Lagrangian},
following the approach presented in \cite{Ivanov:2008er,Ginzburg:2009dp,Ginzburg:2010wa}.  In the   first  approximation the Yukawa couplings and  the quartic coefficients of $V$ are constant, while the quadratic parameters $m_{ii}^2 \; (i=1,2)$ vary with temperature $T$ as follows \cite{Ginzburg:2010wa,grzesiek}:
\bear{c}
m_{ii}^2(T)=  m_{ii}^2-c_iT^2,\\[3mm]
c_1=\frac{3\lambda_1+2\lambda_3+\lambda_4}{6}+\frac{3g^2+g^{\prime 2}}{8}+\frac{g_t^2+g_b^2}{2}, \quad
c_2=\frac{3\lambda_2+2\lambda_3+\lambda_4}{6}+\frac{3g^2+g^{\prime 2}}{8}.
			 \eear{Tempdep}

In this work we limit ourselves to positive $c_1,c_2$ as we consider only the restoration of EW symmetry for high $T$ (the negative values of $m_{11}^2(T)$ and $m_{22}^2(T)$ for high enough $T$). 

As the Universe is cooling down the potential $V$ (\ref{pot}), with temperature dependent quadratic coefficients (\ref{Tempdep}), may have different ground states \cite{Ginzburg:2010wa}. The general form of the neutral extremum is:
\bear{c}
        \langle\Phi_S\rangle =\dfrac{1}{\sqrt{2}}\left(\begin{array}{c} 0\\
        v_S\end{array}\right),\quad \langle\Phi_D\rangle
        =\dfrac{1}{\sqrt{2}}\left(\begin{array}{c} 0 \\ v_D
        \end{array}\right), \quad \left( v^2=v_S^2+v_D^2 \right).
\eear{genvac}

\textit{EW symmetric extremum} ($EW\! s$) is realized if $v_D=v_S=0$. Here all fermions and bosons are massless and $EW$ symmetry is conserved. 

\textit{Inert extremum} $I_1$ can be realized if $v_D=0,\,v_S^2=v^2=m_{11}^2/\lambda_1$. If $I_1$ is a vacuum then in the scalar sector there are four dark scalar particles $D_H,\,D_A,
D^\pm$ and the SM-like Higgs particle $h_S$. The lightest dark particle is stable and so it is a good  DM candidate. Assuming that DM particles are neutral,
we consider such  variant of IDM in which $D_H$ is a DM candidate, meaning $M_{D^\pm}, \, M_{D_A} > M_{D_H}$. Various  theoretical and experimental constraints apply for the IDM (see e.g.~\cite{Cao:2007rm, Dolle:2009fn, LopezHonorez:2006gr, Honorez:2010re, Lundstrom:2008ai, Krawczyk:2009fb,LopezHonorez:2010tb}). EWPT and collider data (LEP II, Tevatron, LHC) constrain the allowed regions of the masses of $h_S$ and dark scalars. The relic density measurements and the direct detection experiments can be used to constrain the DM mass and the DM-Higgs self-coupling $\lambda_{345}$. However, they don't constrain the DM quartic self-coupling $\lambda_2$.

\textit{Inertlike extremum} $I_2$ is mirror-symmetric to $I_1$ as $v_S=0, \, v_D^2=v^2=m_{22}^2/\lambda_2$. Fermions are massless at the tree-level (Model I), while gauge bosons are massive. There are four scalars $S_H,S_A,S^\pm$ (no DM candidate as $D$-symmetry is spontaneously violated) and the Higgs particle $h_D$ with no interaction with fermions.

\textit{Mixed extremum} $M$ is a standard 2HDM extremum with $v_S, v_D \not = 0$. Fermions and bosons are massive and there are 5 Higgs particles: CP-even $h$ and $H$, CP-odd $A$ and charged $H^\pm$, none of them can be a DM candidate.

We assume that today inert phase $I_1$ with the DM candidate $D_H$ is realized. However, the sequences of transitions between different vacua (called here \textit{rays}) were possible in the past \cite{Ginzburg:2010wa}.

There are three types of sequences that start in the $EW\! s$ symmetric phase and end in the inert phase. First one is a single phase transition $EW\! s \to I_1$. It is realized by rays Ia,b,c (for $R>1$, $0<R<1$ and $-1<R<1$, respectively), rays IIa,b ($R>1$ and $0<R<1$) and ray III (only for $R>1$). The difference between those rays is the status of the $I_2$ extremum: for ray I it's not an extremum; for ray II it is an extremum, but not a minimum; for ray III it is a local minimum, but not the global one.

Second type of sequence, $EW\! s \to I_2 \to I_1$, can be realized only if $R>1$ and is represented by the  rays IV and V. In this case the EWSB is a phase transition of a 2nd-order, while the last transition $I_2 \to I_1$ is of the 1st-order. 

Only for $R>1$ there is an unique opportunity of coexistence of minima (vacuum $I_1$ and local metastable minimum $I_2$) for rays III, IV and V. For ray IV the coexistence is temporary and the local minimum $I_2$ disappears for low temperatures, while for rays III and V it still exists for $T=0$.

In the $0<R<1$ case there is a possibility of having three phase transitionss in the sequence $EW\!  s \to I_2 \to M \to I_1$ (ray VI). All transitions here are of the 2nd-order.

\section{Phenomenological analysis}

For phenomenological studies it is useful to chose the physical masses $M_{h_S}$, $M_{D_H}$, $M_{D_A}$, $M_{D^\pm}$ and the scalar self-couplings as an input parameters. In this analysis we chose two self-couplings, $\lambda_{345}$ and $\lambda_2$, which have different properties and play different role in the analysis \cite{Sokolowska:2011sb}.

$\lambda_{345}$ is a triple and quartic coupling of the DM partricle and SM-like Higgs, i.e. $D_H D_H h_S$ or $D_H D_H h_S h_S$. In wide range of DM mass this parameter governs the main annihilation channel into pair of fermions via Higgs exchange: $D_H D_H \to h_S \to f \bar{f}$ with the cross-section $\sigma \propto \lambda_{345}^2/(4M_{D_H}^2 - M_{h_S}^2)^2$. For this reason this parameter, along with the DM mass, influences strongly the value of the DM relic density $\Omega_{DM} h^2$. It also plays an important role in the direct detections experiments, as DM-nucleon elastic scattering cross-section is given by $\sigma_{DM,N} \propto \lambda_{345}^2 /(M_{D_H} + M_{N})^2$ \cite{Barbieri:2006dq}.

The remaining self-coupling, $\lambda_2$, is a quartic coupling of DM particle. For this reason the exact value of $\lambda_2$ does not influence $\Omega_{DM} h^2$ directly. However, this parameter limits $\lambda_{345}$ through the positivity constraints and is important for the type of evolution.

Below we present the analysis done in the $(\lambda_{345},\lambda_2)$ phase space for the three regions of DM mass for chosen benchmark points:
\begin{enumerate}
 \item $M_{D_H} = 5 \textrm{ GeV}, M_{D_A} = 105 \textrm{ GeV}, M_{D^\pm} = 110 \textrm{ GeV}, M_{h_S} = 120 \textrm{ GeV}$,
\item $M_{D_H} = 50 \textrm{ GeV}, M_{D_A} = 120 \textrm{ GeV},M_{D^\pm} = 120 \textrm{ GeV}, M_{h_S} = 120 \textrm{ GeV}$,
\item $M_{D_H} = 800 \textrm{ GeV}, M_{D_A} = 801 \textrm{ GeV}, M_{D^\pm} = 801 \textrm{ GeV}, M_{h_S} = 120\textrm{ GeV}$.
\end{enumerate}

\begin{figure}[t]
\vspace{-10pt}
  \centering
  \subfloat[low DM mass]{\label{low}\includegraphics[width=0.3\textwidth]{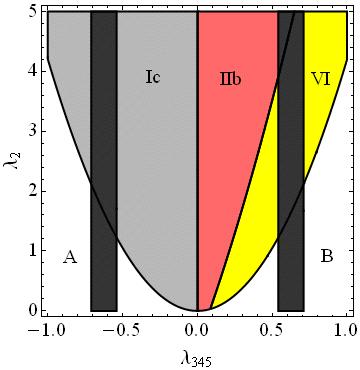}} \quad
  \subfloat[medium DM mass]{\label{mid}\includegraphics[width=0.31\textwidth]{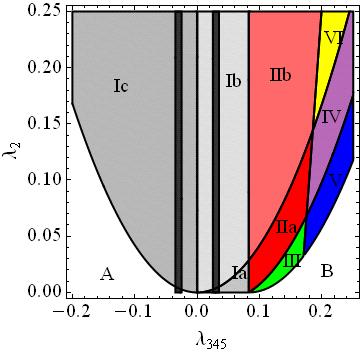}} \quad
  \subfloat[high DM mass]{\label{high}\includegraphics[width=0.3\textwidth]{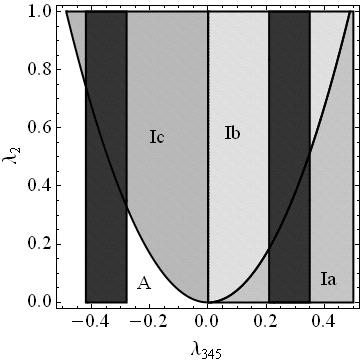}}
  \caption{  Possible rays for different regions of DM mass. Vertical bounds denote region allowed by WMAP measurements; region $A$ is exluded by positivity constraints, in region $B$ $I_1$ is only a \textit{local} minimum. 
}
  \label{fig:rays}
\end{figure}

 Figures \ref{low},\ref{mid},\ref{high} show the possible rays for each region, as well as the 3$\sigma$ WMAP-allowed regions $0.085 < \Omega_{DM} h^2 < 0.139$ \cite{PDG}. Note that the region $A$ is excluded by the positivity constraints and in the region $B$ $I_1$ is only a \textit{local minimum} and not the vacuum. Each ray is realized in the separate region of $(\lambda_{345},\lambda_2)$ phase space. Furthermore, different types of evolution are possible in the cases of low, medium and high DM mass.
 
 \subsection{Low DM mass}
 
  The low mass region, which resembles the singlet scalar DM, has been shown to fit into the CoGeNT, DAMA/Libra and CRESST-II signal \cite{Bernabei:2008yi,Savage:2008er,Aalseth:2010vx,Angloher:2011uu,Andreas:2010dz}, however it appears to be excluded by the XENON100 results \cite{Aprile:2011hi}. In this region the DM particle is much lighter than all other scalar particles with $M_{D_H} \approx (4-8) \textrm{ GeV}$ and $M_{D_A} \approx M_{D^\pm} \approx 100 \textrm { GeV}$ \cite{Dolle:2009fn}. Large mass splittings between the $D_H$ and other scalar particles do not allow for the coannihilation. To have the correct WIMP cross-section and proper relic density rather large $\lambda_{345}$ is needed \cite{Sokolowska:2011sb}.

In this region the possible types of evolution are limited to three rays only (ray Ic, IIb, VI, figure \ref{low}) and there is no coexistence of minima.

Notice, that to fit into the WMAP data we need not only large $\lambda_{345}$, but also large $\lambda_2 \approx 1$. The smaller values of $\lambda_2$ are excluded by positivity constraints (region $A$) or $I_2$ vacuum (region $B$). However, larger $\lambda_{345}$ corresponds to the lower temperature of the final phase transition. In this example for the sequence $EW\!s \to I_2 \to M \to I_1$ it occurs at $T_{M \to I_1} = 6$ GeV, so $M_{D_H} \approx T_{M \to I_1}$. The recent analysis \cite{grzesiek} shows that in this case the lowest order of the thermal corrections to $V$ is not sufficient.

\subsection{Medium DM mass}
 
In this region the DM mass is of the order of $M_{D_H} \approx (45 - 160)$ GeV. Mass splitting between $D_H$ and $D^\pm$ should be large: $M_{D^\pm} - M_{D_H} \approx (50 - 90)$ GeV. Constraints for $M_{D_A} - M_{D_H}$ have been derieved in \cite{Lundstrom:2008ai}. If this value is large with $M_{D_A} \approx M_{D^\pm}$ then there is no coannihilation. For $M_{D_A} - M_{D_H} < 8 \textrm{ GeV}$ this effect influences strongly the value of DM relic density \cite{Dolle:2009fn,LopezHonorez:2006gr}.

For medium DM mass, regardless of the exact values of the mass splittings, all rays are possible (figure \ref{mid}). This is the only case when one can have the 1st-order phase transition for rays IV and V -- those rays are not possible for low or high DM mass. 

 In this region $\Omega_{DM} h^2$ is very sensitive to the exact value of $M_{D_H}$ and mass splittings. Therefore, we cannot make a general statement that a certain ray will always give a proper relic density \cite{Sokolowska:2011sb}. However, some properties of this mass region are independent on $M_{D_H}$. For example, complex sequences (rays IV-VI) require rather large $\lambda_{345}$. This however leads to the similar problem as in the low DM mass case: the temperature of the final phase transition is lower and especially for ray V further thermal corrections to the potential are needed.

 \subsection{High DM mass}

As shown in \cite{LopezHonorez:2006gr} in the high mass region the mass of the dark matter particle should be over $500$ GeV, with the small mass splittings between the dark particles. Also the \textit{perturbative unitarity} may give the relevant constraints \cite{bogusia}. To have a proper relict density we need rather large absolute values of $|\lambda_{345}|$ and $\lambda_{2}$ \cite{Sokolowska:2011sb}. Only three rays are possible; they correspond to the sequence with a single phase transition $EW\! s \to I_1$ and differ only by the value of $R$ (rays Ia, Ib, Ic, figure \ref{high}). Other types of evolution require $\lambda \approx O(10-20)$ \cite{Sokolowska:2011sb}.

\subsection{Role of $\lambda_2$ self-coupling}

\begin{figure}[t]
\vspace{-10pt}
  \centering
  \subfloat[ray V, $\lambda_2 = 0.0684$]{\label{r5}\includegraphics[width=0.4\textwidth]{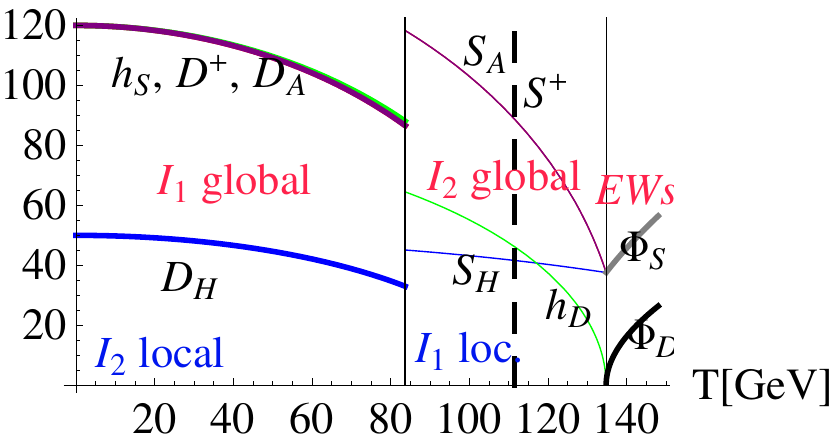}} \quad
  \subfloat[ray VI, $\lambda_2 = 0.1672$]{\label{r6}\includegraphics[width=0.4\textwidth]{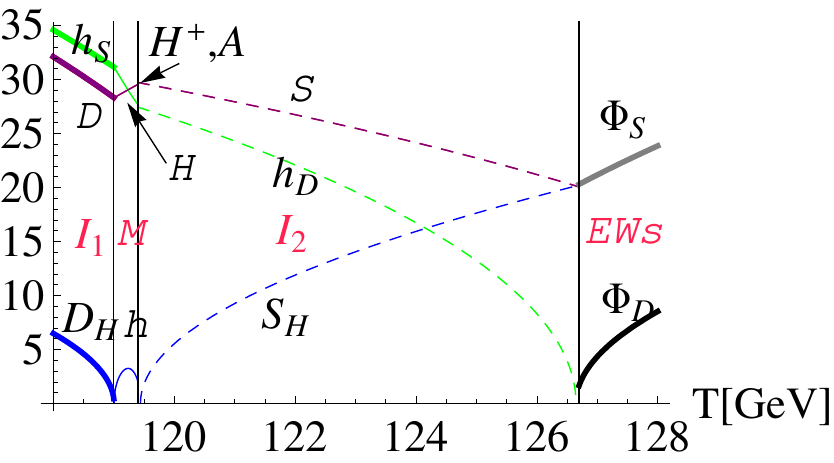}}
  \caption{  Temperature evolution of masses of scalar particles in different sequences. $M_{D_H} = 50 \textrm{ GeV}, M_{D_A} = 120 \textrm{ GeV},M_{D^\pm} = 120 \textrm{ GeV}, M_{h_S} = 120 \textrm{ GeV}, \lambda_{345} = 0.1945$. Notation: $D=D^\pm,\, D_A$, $S=S^\pm,\, S_A$.
}
  \label{fig:lambda}
\end{figure}

$\lambda_2$ self-coupling is a significant parameter in IDM -- as shown in the previous sections. It not only limits $\lambda_{345}$ through the positivity constraints, but also is important for the type of evolution. In this section we fix the scalar masses and $\lambda_{345}$ self-coupling and let $\lambda_2$ vary. Depending on the value of $\lambda_2$, different types of evolution are realized.

Figures \ref{r5},\ref{r6} show the thermal evolution of the mass parameters during evolution of the Universe \cite{Sokolowska:2011yi}. In case of ray V (figure \ref{r5}) there are two phase transitions. EWSB into $I_2$ phase happens for $T = 134.8 \textrm{ GeV}$. Dashed line shows the apperance of the local minimum $I_1$ during the inertlike phase of the evolution. $I_1$ becomes a global minimum after 1st-order phase transtion which takes place for $T=83.7 \textrm{ GeV}$. From this point $I_2$ is a local minimum that still exists for $T=0$. 

Figure \ref{r6} shows the evolution according to ray VI. Here, after EWSB at $T = 126.7\textrm{ GeV}$ Universe enters the inertlike phase with massless
fermions and massive gauge bosons. This minimum becomes a saddle point for $T=119.4 \textrm{ GeV}$ and the 2nd-order transition to the M phase takes place. This phase is short-lived and soon, at $T = 119.0 \textrm{ GeV}$, there is another 2nd-order transition into $I_1$ phase. There is no coexistence of minima at any point in time during evolution.

Notice, that although the temperature of EWSB is similar in both cases, the final phase transition happens at different temperatures and it's much lower for ray V. As discussed, this ray is the most likely to require further thermal corrections to the potential \cite{grzesiek}.

\section{Conclusions}

In this paper we studied the temperature evolution of the Universe in IDM. We also discuss the significance of the quartic self-coupling $\lambda_2$. This parameter does not influence DM relic density directly and it cannot be accessible in the colliders. However, it is related to the value of $\lambda_{345}$ coupling through the positivity constraints. It also plays an important role in the evolution, as its different values lead to the different types of the evolution of the Universe.

\end{document}